\documentclass{mem}
\usepackage{natbib}
\usepackage{txfonts}
\usepackage{balance}
\usepackage{graphicx}
\usepackage[a4paper,breaklinks,dvipdfm]{hyperref}
\idline{75}{282}

\begin{document}

\def\teff{$T\rm_{eff }$}
\def\kms{$\mathrm {km s}^{-1}$}

\title{OJ 287 binary black hole system}

%%\subtitle{}
\author{
M. \,Valtonen\inst{1},
S. \,Ciprini\inst{2}
}

\institute{%
Helsinki Institute of Physics, University of Helsinki,
FIN-00014 University of Helsinki, Finland
and
Finnish Centre for Astronomy with ESO, University of Turku,
FIN-21500 Piikki\"o, Finland
\and
ASI Science Data Center and INAF Observatory of Rome, Via G. Galilei, P.O.Box 64,
00044 Frascati, Italy\\
\email{mvaltonen2001@yahoo.com}
}

\offprints{M.Valtonen}

\authorrunning{Valtonen}

\titlerunning{OJ 287}

\abstract{
The light curve of the quasar OJ~287 extends from 1891 up today without major gaps. This is partly due to extensive studies of historical plate archives by Rene Hudec and associates, partly due to several observing campaigns in recent times. Here we summarize the results of the 2005 - 2010 observing campaign in which several hundred scientists and amateur astronomers took part. The main results are the following: (1) The 2005 October optical outburst came at the expected time, thus confirming the general relativistic precession in the binary black hole system. This result disproved the model of a single black hole system with accretion disk oscillations, as well as several toy models of binaries without relativistic precession. In the latter models the main outburst would have been a year later. (2) The nature of the radiation of the 2005 October outburst was expected to be bremsstrahlung from hot gas at the temperature of $3\times 10^{5}$ $^{\circ}$K. This was confirmed by combined ground based and ultraviolet observations using the XMM-Newton X-ray telescope. (3) A secondary outburst of the same nature was expected at 2007 September 13. Within the accuracy of observations (about 6 hours), it started at the correct time. Thus the prediction was accurate at the same level as the prediction of the return of Halley's comet in 1986. (4) Further synchrotron outbursts were expected following the two bremsstrahlung outbursts. They came as scheduled between 2007 October and 2009 December. (5) Due to the effect of the secondary on the overall direction of the jet, the parsec scale jet was expected to rotate in the sky by a large angle around 2009. This rotation may have been seen at high frequency radio observations. OJ~287 binary black hole system is currently our best laboratory for testing theories of gravitation. Using OJ~287, the correctness of General Relativity has now been demonstrated up to the third Post-Newtonian order, at higher order than has been possible using the binary pulsars.
}
\maketitle{}

\section{Introduction}

In 1982 Aimo Sillanp\"a\"a put together the historical light curve of OJ~287 based on published measurements. There appeared to be a 12 year outburst cycle (see Figure 1), and moreover, it was obvious that the next cyclic outburst was due very shortly. Indeed, OJ~287 produced the expected event in the following January \citep{haarala83,sillanpaa85}. Observations showed a sharp decline in the percentage polarization during the outburst maximum, indicating that the outburst was produced essentially by unpolarized light \citep{smith87}. This is different from ordinary outbursts in OJ~287 which are characterized by an increase of the percentage polarization at the maximum light. Radio outbursts were found to follow the optical outbursts with a time delay of between 2 months and a year, depending on the observing frequency \citep{valtaoja87}.

\begin{figure}
\includegraphics[width=4.5cm,angle=270]{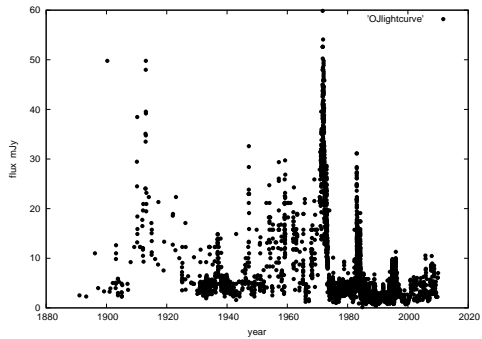}
 \caption{The optical light curve of OJ~287 from 1891 to 2010. It includes unpublished data from R.Hudec and M.Basta}
\end{figure}

\citet{sillanpaa88} suggested that OJ~287 is a binary black hole system where a smaller companion perturbs periodically the accretion disk of a massive primary black hole. The next expected outburst was in 1994; it came as scheduled \citep{fiorucci 94,sillanpaa96a}. In the binary model there should be two disk crossings per 12 yr orbital period. Thus the 1994 outburst should have an equal pair whose timing was calculated to be at the beginning of 1995 October \citep{valtonen96,lehto96,sundelius96}. This prediction was also verified \citep{sillanpaa96b}.

Alternative explanations were also put forward. Quasiperiodic oscillations in an accretion disk were suggested \citep{igumenshchev99}, and several binary toy models without relativistic precession were also proposed \citep{katz97,villata98,valtaoja00}. The latter models all predicted the next main outburst of OJ~287 in the autumn of 2006 while the precessing binary model gave a prediction one year earlier, at the beginning of 2005 October \citep{sundelius97,kidger00}. The second major outburst was expected in late 2007 in all binary models while in the oscillating accretion disk model there was no reason to expect a second major outburst. In the precessing binary model the date was given with high accuracy, with the last prediction prior to the actual event being 2007 September 13 \citep{valtonen07,valtonen08a}. In the oscillating accretion disk model and in non-precessing binary models the nature of the radiation at these outbursts should have been polarized synchrotron radiation while the precessing binary model predicted unpolarized bremsstrahlung radiation \citep{lehto96}. In addition, the precessing binary model predicted a series of further outbursts for the interval 2007 - 2010 \citep{sundelius97}. In this model the companion black hole should also affect the disk of the primary in a predictable way, leading to the wobble of the jet \citep{valtonen06}.

With these predictions in mind, a multiwavelength campaign of observing OJ 287 during 2005 - 2010 was set up.

\section{Five 'smoking gun' results}

In the following we describe five 'smoking gun' observations which produced expected results from the point of view of the precessing binary black hole model but which were surprising and unexpected in other theories.

\subsection{Timing the 2005 outburst}

The 2005 outburst was well covered by altogether 2329 observations in V and R-bands. The points in Figure 2 are daily averages, 92 in all.

\begin{figure}
\includegraphics[width=4.5cm,angle=270]{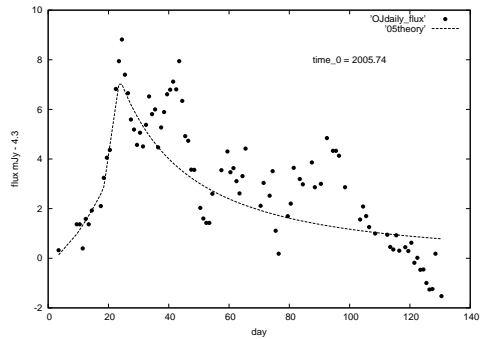}
 \caption{The optical light curve of OJ~287 during the 2005 outburst. The dashed line is the theoretical fit based on \citep{lehto96}.}
\end{figure}

According to \citep{sundelius96}, the impact causing the 2005 outburst was expected 22.3 years after the impact of the 1983 outburst. In addition, \citep{lehto96} estimated that the 2005 outburst should be delayed after the impact. Also the 1983 outburst is delayed but not as much, the difference being 0.44 yr. The rapid flux rise started in the latter outburst at 1983.00; thus the corresponding rapid flux rise of the 2005 outburst was expected at 1983.00 + 22.30 + 0.44 = 2005.74. Actually, the outburst was one week late and did not begin until 2005.76 \citep{valtonen08c}, but anyway the timing was well within the error limits. Only a few polarization measurements were carried out at that time, and unfortunately, even those happened during secondary flares. Thus the polarization state of the primary outburst is not known.

The timing confirmed the precession of $39.1^{\circ}$ per period. It is so much higher that e.g. in  binary pulsars (by a factor of $10^4$) that we immediately realise the importance of OJ~287 in testing General Relativity.

\begin{figure}
\includegraphics[width=4.5cm,angle=270]{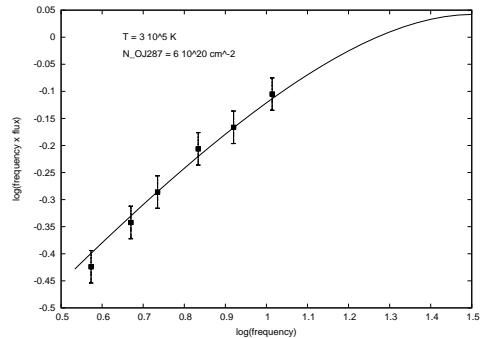}
 \caption{The optical - UV spectrum of OJ~287 during the 2005 outburst. The solid line is the bremsstrahlung fit as predicted in \citep{lehto96}. The observational points are corrected for the internal extinction in OJ~287 as well as for the extinction in our Galaxy.}
\end{figure}

\subsection{Nature of radiation at  the 2005 outburst}

The impact outbursts are expected to consist of bremsstrahlung radiation and thus the optical polarization of OJ~287 should go down during them. As mentioned above, the polarization information for the basic 2005 outburst is not available. However, bremsstrahlung may be recognized also by it spectrum, and this is the part of the campaign that was successfully carried out.

\begin{figure}
\includegraphics[width=4.5cm,angle=270]{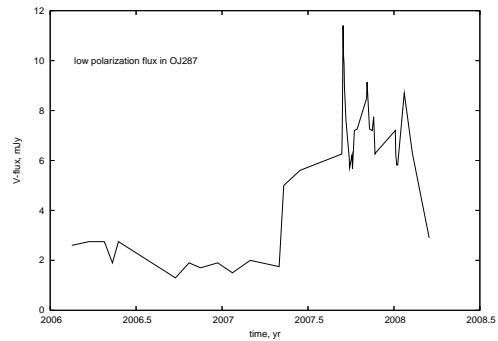}
 \caption{The optical light curve of OJ~287 during 2006-2008. Only low polarization (less than $10\%$) data points are shown.}
\end{figure}

We had XMM-Newton observations both before the 2005 outburst (2005 April), and during the outburst (2005 November 3-4). Fortunately the November observation happened at the time when the source was at its basic outburst level, in between two secondary bursts. Thus we would expect to see an additional pure bremsstrahlung spectrum above the underlying synchrotron power-law. A preliminary report of these observations has appeared in \citep{ciprini07}, and a more detailed report is under preparation.

\begin{figure}
\includegraphics[width=4.5cm,angle=270]{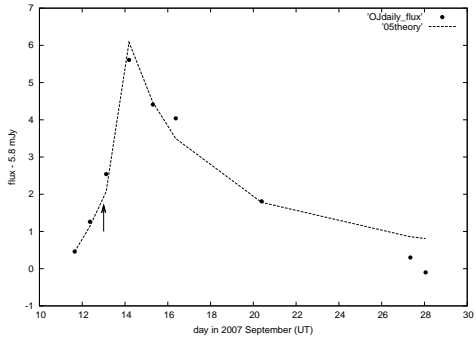}
 \caption{The optical light curve of OJ~287 during the 2007 September outburst. Only low polarization (less than $10\%$) data points are shown. Data points are based on Ref. 22. The dashed line is the theoretical fit based on Ref 9. The arrow points to September 13.0, the predicted time of origin of the rapid flux rise.}
\end{figure}

In Figure 3 we show the difference between the 2005 November flux and the 2005 April flux. The values have been corrected for the Galactic extinction and for the extinction in the OJ~287 host galaxy. For the latter, we use the measurements in \citet{gosh95} and the standard Galactic extinction curve \citep{draine03}. The solid line shows the bremsstrahlung spectrum at the expected temperature of $3\times 10^{5}$ $^{\circ}$K. Note that a raised synchrotron spectrum, as one might have expected in some other theories, would have a downward slope toward higher frequencies, and it is entirely inconsistent with observations.

\subsection{Timing and nature of the 2007 September 13 outburst}

The 2007 September 13 outburst was an observational challenge, as the source was visible only for a short period of time in the morning sky just before the sunrise. Therefore a coordinated effort was made starting with observations in Japan, then moving to China, and finally to central and western Europe.

Four outbursts of various sizes were detected in 2007 September. In three of them the degree of polarization was above $15\%$ while the fourth and the biggest one had polarization below $10\%$ \citep{valtonen08b}. Thus it is not difficult to decide which one was the expected bremsstrahlung event. Later in the year there were more highly polarized outbursts, but if we look at the light curve composed of low polarization states only (Figure 4), the September 13 outburst clearly stands out as a sharp spike. The light curve has been plotted with a better resolution in Figure 5. A theoretical light curve is also drawn, and an arrow points to the expected moment of the beginning of the sharp flux rise. We see that the observed flux rise coincides within 6 hours with the expected time. The accuracy is about the same with which we were able to predict the return of Halley's comet in 1986!

\subsection{2007 - 2010 outbursts}

\citet{sundelius97} gave a detailed prediction of the whole light curve of OJ~287 during the campaign period. In addition to the two impact outbursts, it was expected that the tidal forcing mechanism of \citep{sillanpaa88} would create a long hump in the light curve, starting in the middle 2007 and continuing until the middle of 2009. This hump is clearly seen in Figure 5, together with the 2007 September 13 spike. \citet{seta09} carried out multifrequency observations of OJ~287 at the hump (2007 November 7-9) and compared the flux values with earlier observations before the hump (2007 April). They find that the excess radiation at the hump is neither due to increased magnetic field strength nor increased Lorentz factor, but likely results from an increase in electron energy density. This is what is expected in the tidal model.

A prominent outburst was also expected at the end of 2009 and it was observed \citep{valtonen11a}.

\subsection{Jet reorientation}

The accretion disk as a whole is also affected by the companion in its 12 year orbit. If the jet and the disk are connected the jet direction should be influenced by the companion.

There are three periodicities that could be expected to show up: the 12 yr orbital cycle, the 120 yr precession cycle (or half of it due to symmetry) and the Kozai cycle \citep{innanen97} which also happens to be $\sim120$ yr. The 12 yr orbital cycle produces the tidal enhancements in accretion flow, and the enhancement can be stronger or weaker depending on where we are in the precession cycle. These two tidal effects explain the overall appearance of the light curve \citep{sundelius97}. In addition, there is a modulation in the long term base emission level due to the jet wobble which follows the Kozai cycle.

The jet wobble shows up in observations in several ways. First, the mean angle of optical polarization varies \citep{sillanpaa91}. The binary model predicts, among other things, a quick change in the optical polarization angle by nearly $90^{\circ}$ around 1995 which has been observed \citep{villforth10}. In radio, we should see a similar rapid change in the position angle of the parsec scale jet. Depending on the delay in radio jet reorientation, the change could be already under way (Figure 6), or it may be delayed by another 12 yr cycle \citep{valtonen11b}. There are recent observations which suggest the first alternative \citep{agudo11}, but the interpretation of these observations is not yet clear-cut.

\begin{figure}
\includegraphics[width=4.5cm,angle=270]{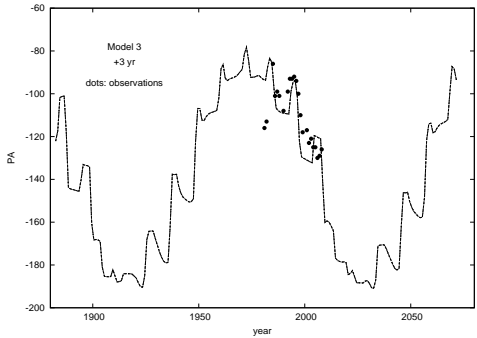}
 \caption{The observed position angle of the radio jet of OJ~287 (points) compared with the binary model, with a 3 year jet response time to the reorientation of the disk.}
\end{figure}

\section{Testing General Relativity}

Using the OJ~287 binary, we may test the idea that the central body is actually a black hole. One of the most important characteristics of a black hole is that it must satisfy the so called no-hair theorem \citep{misner73}. A  practical test was suggested by \citet{thorne80}. In this test the quadrupole moment $Q$ of the spinning body is measured. If the spin of the body is $S$ and its mass is $M$, we determine the value of $q$ in

\begin{eqnarray}
Q = -q \frac{S^2}{Mc^2}.
\end{eqnarray}

For black holes $q=1$, for neutron stars and other possible bosonic structures $q > 2$ \citep{wex99,will08}

We have studied the distribution of $q$-values in orbits which give the correct timing of 9 outbursts within the range of measurement accuracy. We find that the distribution peaks at $q=1$, thus confirming the no-hair theorem. The parameter values $q=0$ or $q=2$ can be rejected at the 3 standard deviation level at present. This is the first time that it has been possible to study general relativity at higher than the 1.5PN order.

\section{Conclusions}

Prior to the 2005 - 2010 multiwavelength campaign there were several ideas about the nature of OJ~287. Fortunately, these models made completely different predictions about the behaviour of OJ~287. The result of a scientific enquiry is seldom as clear-cut as this: the outbursts satisfied the expectations of the precessing binary model in every respect while the alternative ideas failed. Therefore we are confident that OJ~287 can be used as a test laboratory for theories of gravitation. Our first results strongly support general relativity as the correct theory.

\begin{acknowledgements}
We would like to thank all the participants of this campaign for an extraordinary amount of work dedicated to the solving of the riddle of OJ~287.
\end{acknowledgements}

\bibliographystyle{aa}

\bigskip

\noindent {\bf DISCUSSION}

\bigskip
\noindent {\bf FRANK RIEGER:} (1) What is the gravitational lifetime of the system? I presume it is short. (2) You seem to need that the secondary crosses the primary disk twice per orbit. I remember that \citep{ivanov99} found that the co-alignment (orbital plane vs. disk plane) happens relatively quickly. Did you check for that?

\bigskip
\noindent {\bf MAURI VALTONEN:} (1) 10000 yr. (2) Observational evidence is that the disk and the orbital plane are at roughly 90 degree angle relative to each other. This comes primarily from the so called ``eclipses'' which may arise when the secondary crosses the jet. Theoretically, the first order perturbation theory shows that the inclination can remain at 90 degrees if it has obtained a value close to it at some earlier time. Our numerical simulations confirm this.

\bigskip
\noindent {\bf MANEL PERUCHO:} There is tendency to associate the radio jet helical structures to precession caused by binary black hole systems. Could comment on this point? Is the periodicity you find in the jet of OJ~287 the expected one from your model?

\bigskip
\noindent {\bf MAURI VALTONEN:} \citet{marscher11} have discovered that in the large scale X-ray and radio jet there are oscillations in the time scale of thousands of years. In our model the spin axis of the primary precesses with the period of about 1300 years \citep{valtonen10}. These two facts may be connected.

\end{document}